\pdfoutput=1
\documentclass[11pt]{article}
\usepackage[T1]{fontenc}
\usepackage[utf8]{inputenc}
\usepackage[margin=1in]{geometry}
\usepackage{amsmath,amssymb,bm}
\usepackage{graphicx}
\usepackage{booktabs}
\usepackage{microtype}
\usepackage{authblk}
\usepackage{enumitem}
\usepackage{hyperref}
\usepackage{xcolor}
\usepackage{caption}
\usepackage{setspace}
\hypersetup{colorlinks=true,linkcolor=blue,citecolor=blue,urlcolor=blue}
\newcommand{\FC}{F_{\!C}}
\newcommand{\FQ}{F_{\!Q}}
\newcommand{\LB}{\mathcal{L}}
\newcommand{\E}{\mathbb{E}}
\newcommand{\Var}{\operatorname{Var}}

\title{\textbf{Fisher-Information Certification from Bell Statistics: CHSH Bounds, Information Loss, and Finite-Data Guarantees}}
\author[1,*]{Juan Sumaya-Mart\'inez}
\author[2]{Omar Olmos-L\'opez}
\author[1]{Jorge Mulia Rodriguez}
\affil[1]{Faculty of Sciences, Universidad Aut\'onoma del Estado de M\'exico, Toluca 50000, Mexico; emails: jsm@uaemex.mx (J.S.-M.), jmr@uaemex.mx (J.M.R.)}
\affil[2]{Tecnol\'ogico de Monterrey, Campus Estado de M\'exico, Estado de M\'exico, Mexico; email: oolmos@tec.mx}
\affil[*]{Corresponding author: jsm@uaemex.mx}
\date{}

\begin{document}
\maketitle


\begin{abstract}
Bell-test records contain more metrological information than the single Bell parameter usually reported. This work quantifies how much parameter sensitivity remains after compressing a Bell experiment to correlators or to a scalar Bell score. A standard information inequality is specialized to differentiable full-correlation Bell statistics, yielding a lower bound on the classical Fisher information of the observed record. For uniformly sampled CHSH settings, the bound becomes $\FC\ge S'^{2}/(16-S^{2})$ and remains valid with biased marginals. For balanced marginals, retaining the four correlators gives a tighter certificate and an exact expression for the Fisher information of the chosen measurement, establishing an explicit information hierarchy. Applied to entangled-photon retardance sensing, the hierarchy reveals a Bell--metrology tradeoff: maximal CHSH violation occurs where the first-order Bell-score certificate vanishes even though the full coincidence record remains informative. The analysis also treats accidental and multipair contamination and provides a nonasymptotic finite-count lower-confidence certificate. The resulting framework converts Bell-calibration data into a controlled quantum-sensing statement while keeping nonlocality, sensitivity, and metrological advantage operationally distinct.
\end{abstract}

\section{Introduction}

Bell tests and parameter-estimation experiments often use much of the same laboratory hardware, but they ask different questions. A Bell inequality tests whether a set of observed correlations is compatible with a local-hidden-variable model \cite{CHSH1969}. Quantum metrology asks how strongly observed probabilities respond to a parameter and how precisely that parameter can be estimated \cite{BraunsteinCaves1994,Paris2009,PezzeRMP2018}. The classical Fisher information (FI) describes a fixed measurement; the quantum Fisher information (QFI) is a state-level optimization over measurements. These notions are related through the probability model, but they are not interchangeable resources. In particular, a large QFI does not in general imply Bell nonlocality \cite{Frowis2019}.

Several nearby lines of work make the distinction especially important. QFI can be witnessed from incomplete sets of observables without full tomography \cite{Apellaniz2017}, and moment hierarchies can lower-bound the FI carried by imperfect measurements \cite{Len2022}. Bell nonlocality has also been connected to quantum-enhanced metrology in many-body systems, where Bell correlators can witness states with large QFI \cite{Niezgoda2021}. In polarization optics, entangled two-channel measurements have been shown to improve polarimetric precision under appropriate resource and noise assumptions \cite{Pedram2024}. More broadly, finite Bell data can be used for device-independent property estimation \cite{Lin2018}, while recent measurement-device-independent certification work illustrates the active interface between metrology and quantum certification \cite{Larsen2026}.

Our question is different from each of these. We do not attempt to infer the QFI of an unknown state from Bell violation, nor do we identify nonlocality with a metrological resource. Instead, we ask how much \emph{classical} Fisher information is guaranteed in the Bell-test record itself after that record is compressed to correlators or to a single Bell score. This shift from state-level QFI witnessing to record-level FI certification is the operational point of the present work.

A Bell value by itself cannot answer that question. Fisher information depends on the full parameter-dependent probability distribution, whereas the Bell parameter is only one linear combination of correlators. The relevant quantity is therefore not just the value of a Bell score but its local response to the parameter. The first-moment information inequality used below is standard \cite{Cramer1946,Len2022}; our contribution is its Bell-statistics specialization, the resulting hierarchy of information loss, the CHSH Bell--metrology tradeoff, and a finite-data certification procedure tied directly to the observed coincidence record.

For an arbitrary differentiable full-correlation Bell score, we construct an event-level random variable whose expectation is the measured Bell score. The covariance between that random variable and the likelihood score gives a lower bound on the FI. For uniformly sampled CHSH settings the result is
\begin{equation}
\FC(\phi)\ge \frac{S'(\phi)^2}{16-S(\phi)^2}.
\label{eq:universalCHSH}
\end{equation}
This bound does not require balanced single-party marginals. If the marginals are balanced, the four correlators can be retained separately and provide both an exact fixed-measurement FI and a tighter lower bound. The resulting hierarchy makes explicit how much metrological information is lost when a full coincidence record is compressed to four correlators and then to one Bell number.

The polarization example also resolves a point that can otherwise be misleading. With canonical CHSH analyzers, the complete coincidence record can be QFI-optimal for the ideal retardance model. Yet at the point of maximum CHSH violation the Bell score is stationary, so a certificate based only on its first derivative is zero. Bell-optimal and Bell-score-metrology-optimal operating points are therefore not the same. We make this tradeoff explicit and compare the result with an optimized separable probe, which reaches the same QFI for the local phase encoding considered here.

The rest of the paper addresses the experimental issues that matter for using the bound in practice. We keep biased marginals in the full likelihood, model visibility loss and accidental/multipair contamination, compare the information retained by Bell statistics with what is available from full state tomography, and construct a finite-count lower-confidence certificate on a single simultaneous probability event. The aim is not to identify Bell violation with metrological advantage, but to show precisely what a Bell-calibration data stream can and cannot certify.

\section{Bell experiment as a parametric statistical model}

Alice and Bob choose settings $x\in\{0,1\}$ and $y\in\{0,1\}$ with probabilities $p_{xy}>0$, $\sum_{xy}p_{xy}=1$, and record dichotomic outcomes $a,b\in\{\pm1\}$. The unknown parameter is $\phi$. The complete single-trial distribution is
\begin{equation}
q_\phi(x,y,a,b)=p_{xy}P_\phi(a,b|x,y),
\label{eq:joint}
\end{equation}
where throughout we assume that the random setting probabilities do not depend on $\phi$. The classical FI per detected event is
\begin{equation}
\FC(\phi)=\sum_{xyab}q_\phi(x,y,a,b)
\left[\partial_\phi\ln q_\phi(x,y,a,b)\right]^2.
\label{eq:FIjoint}
\end{equation}
Because $p_{xy}$ is parameter independent, Eq.~(\ref{eq:FIjoint}) is equivalently $\sum_{xy}p_{xy}F_{xy}$, where $F_{xy}$ is the FI of the conditional outcome distribution for the setting pair $(x,y)$.

Define the full correlators
\begin{equation}
E_{xy}(\phi)=\sum_{ab}ab\,P_\phi(a,b|x,y),
\end{equation}
and a general full-correlation Bell score
\begin{equation}
B(\phi)=\sum_{xy}c_{xy}E_{xy}(\phi).
\label{eq:BellScore}
\end{equation}
CHSH corresponds to $c_{00}=c_{01}=c_{10}=+1$ and $c_{11}=-1$, in which case $B=S$.

No assumption about vanishing marginals has yet been made. The most general binary-outcome distribution can be written
\begin{equation}
P_\phi(a,b|x,y)=\frac{1}{4}\left[1+a m_x+b n_y+abE_{xy}\right],
\label{eq:biasedP}
\end{equation}
where all three quantities may depend on $\phi$. The positivity of the four probabilities imposes the usual constraints on $(m_x,n_y,E_{xy})$.

\subsection{Scope, trust assumptions, and postselection}

The bounds below are statements about the explicitly specified observed record. We assume a differentiable family of distributions with parameter-independent support, parameter-independent setting probabilities, and trials that are sufficiently stationary for the single-trial likelihood to describe the data (independent trials are used for the finite-count construction). The analysis is therefore not a loophole-free or fully device-independent Bell certification. In particular, if the experimental record is conditioned on detected coincidences, the Fisher information is 
\emph{per retained coincidence}. A resource comparison per emitted pair or per unit time must instead include no-click outcomes and any parameter dependence of the total detection probability. Likewise, a CHSH value calculated after fair-sampling postselection should be described as a Bell-type statistic unless the detection loophole is independently closed. These distinctions do not affect the statistical inequality itself, but they are essential when interpreting it as nonlocality or as a metrological resource statement.

\section{A Bell-score lower bound on Fisher information}

\subsection{General information inequality}

We first prove the central result without using Eq.~(\ref{eq:biasedP}). Define, for one experimental event,
\begin{equation}
W(x,y,a,b)=\frac{c_{xy}ab}{p_{xy}}.
\label{eq:W}
\end{equation}
Its expectation under Eq.~(\ref{eq:joint}) is exactly the Bell score,
\begin{equation}
\E_\phi[W]=B(\phi).
\end{equation}
Let $\ell_\phi=\partial_\phi\ln q_\phi$ denote the likelihood score. Under the standard regularity condition $\E_\phi[\ell_\phi]=0$,
\begin{equation}
B'(\phi)=\partial_\phi\E_\phi[W]=\E_\phi[W\ell_\phi]
=\operatorname{Cov}_\phi(W,\ell_\phi).
\end{equation}
Cauchy--Schwarz therefore gives
\begin{equation}
[B'(\phi)]^2\le \Var_\phi(W)\,\FC(\phi).
\label{eq:infoineq}
\end{equation}
Since $a^2b^2=1$,
\begin{equation}
\E[W^2]=\sum_{xy}\frac{c_{xy}^2}{p_{xy}},
\end{equation}
so that
\begin{equation}
\boxed{
\FC(\phi)\ge
\LB_B(\phi)\equiv
\frac{[B'(\phi)]^2}{\displaystyle\sum_{xy}c_{xy}^2/p_{xy}-B(\phi)^2}
}
\label{eq:generalBellBound}
\end{equation}
whenever the denominator is nonzero. Degenerate zero-variance cases are understood by continuity; then regularity also forces the corresponding first derivative to vanish.

Equation~(\ref{eq:generalBellBound}) is a classical statistical statement about the observed record. It does not infer the complete FI from a Bell value; instead it lower-bounds the complete FI using the local response of a particular statistic. It requires neither a qubit state nor unbiased marginals, and it remains valid for mixed states, detector imbalance, and static background counts as long as these effects are incorporated in $P_\phi$.

For CHSH with uniform settings $p_{xy}=1/4$,
\begin{equation}
\boxed{
\FC(\phi)\ge \LB_S(\phi)
=\frac{S'(\phi)^2}{16-S(\phi)^2}.
}
\label{eq:scoreCHSH}
\end{equation}
This is the weakest but most assumption-robust certificate considered below.

\subsection{Optimal setting allocation for score-only certification}

The setting probabilities can themselves be optimized. For a fixed Bell coefficient vector, the only $p_{xy}$-dependent term in Eq.~(\ref{eq:generalBellBound}) is
\begin{equation}
A(p)=\sum_{xy}\frac{c_{xy}^2}{p_{xy}}.
\end{equation}
Minimizing $A$ subject to $\sum p_{xy}=1$ gives
\begin{equation}
p_{xy}^{\star}=\frac{|c_{xy}|}{\sum_{x'y'}|c_{x'y'}|},
\label{eq:popt}
\end{equation}
with
\begin{equation}
A_{\min}=\left(\sum_{xy}|c_{xy}|\right)^2.
\end{equation}
Thus equal sampling is optimal for the score-only CHSH certificate because all four CHSH coefficients have equal magnitude. For nonuniform Bell witnesses, Eq.~(\ref{eq:popt}) gives the natural event-allocation rule.

\section{What extra information is carried by the individual correlators?}

\subsection{Exact full-probability FI with biased marginals}

Equation~(\ref{eq:generalBellBound}) is valid whether or not the marginals are balanced. If the full probabilities are recorded, one need not discard the corresponding information. For Eq.~(\ref{eq:biasedP}), the setting-resolved FI is exactly
\begin{equation}
F_{xy}=\frac{1}{4}\sum_{a,b=\pm1}
\frac{\left[a m'_x+b n'_y+ab E'_{xy}\right]^2}
{1+a m_x+b n_y+abE_{xy}},
\label{eq:fullbiasedFI}
\end{equation}
and $\FC=\sum_{xy}p_{xy}F_{xy}$. Equation~(\ref{eq:fullbiasedFI}) is the correct expression when unequal detector efficiencies, polarization-dependent loss, imperfect splitting, or asymmetric accidental backgrounds generate nonzero marginals. No symmetrization is mathematically required if these probabilities are measured reliably.

\subsection{Unbiased-marginal identity and tighter certificate}

If $m_x=n_y=0$, then
\begin{equation}
P(a,b|x,y,\phi)=\frac{1+abE_{xy}(\phi)}{4},
\end{equation}
and direct substitution in Eq.~(\ref{eq:FIjoint}) gives
\begin{equation}
\boxed{
\FC(\phi)=\sum_{xy}p_{xy}
\frac{[E'_{xy}(\phi)]^2}{1-E_{xy}(\phi)^2}.
}
\label{eq:exactCorrFI}
\end{equation}
This is an exact FI of the observed CHSH measurement, not a QFI and not a reconstruction of the state.

Applying Cauchy--Schwarz to $B'=\sum c_{xy}E'_{xy}$ yields the correlator-resolved lower bound
\begin{equation}
\boxed{
\FC(\phi)\ge \LB_E(\phi)\equiv
\frac{[B'(\phi)]^2}
{\displaystyle\sum_{xy}\frac{c_{xy}^2[1-E_{xy}(\phi)^2]}{p_{xy}}}.
}
\label{eq:corrBound}
\end{equation}
For CHSH with uniform settings,
\begin{equation}
\LB_E=\frac{S'^2}{4\sum_{xy}[1-E_{xy}^2]}.
\end{equation}
Moreover, $\LB_E\ge \LB_B$. Indeed,
\begin{align}
&\left[\sum_{xy}\frac{c_{xy}^2}{p_{xy}}-B^2\right]
-\sum_{xy}\frac{c_{xy}^2(1-E_{xy}^2)}{p_{xy}}\nonumber\\
&=\sum_{xy}\frac{c_{xy}^2E_{xy}^2}{p_{xy}}-\left(\sum_{xy}c_{xy}E_{xy}\right)^2\ge0,
\end{align}
where the last inequality follows from Cauchy--Schwarz and $\sum p_{xy}=1$.

We therefore obtain the operational hierarchy
\begin{equation}
\boxed{\qquad
\FC\ \ge\ \LB_E\ \ge\ \LB_B,
\qquad}
\label{eq:hierarchy}
\end{equation}
where $\LB_E$ requires unbiased marginals but $\LB_B$ does not. The hierarchy quantifies the price paid for compressing a multi-outcome data record first to four correlators and then to a single Bell score.

\section{Polarization-entangled retardance model}

Let $|H\rangle\equiv|0\rangle$, $|V\rangle\equiv|1\rangle$ and
\begin{equation}
|\Phi^+\rangle=\frac{|HH\rangle+|VV\rangle}{\sqrt2}.
\end{equation}
A local retarder in Alice's arm implements
\begin{equation}
U_A(\phi)=\exp(-i\phi\sigma_z/2),\qquad
|\psi_\phi\rangle=(U_A\otimes I)|\Phi^+\rangle.
\end{equation}
For equatorial observables
\begin{equation}
\sigma_\theta=\cos\theta\,\sigma_x+\sin\theta\,\sigma_y,
\end{equation}
we use opposite phase conventions on the two arms,
\begin{equation}
A_x=\sigma_{-\alpha_x},\qquad B_y=\sigma_{\beta_y},
\end{equation}
so that
\begin{equation}
E_{xy}(\phi)=\cos(\alpha_x-\beta_y+\phi).
\label{eq:Eideal}
\end{equation}
This sign convention is only a labeling choice for the equatorial analyzer phases; an explicit derivation is given in Appendix~\ref{app:qubitderivation}.
The canonical CHSH angles
\begin{equation}
\alpha_0=0,\quad \alpha_1=\frac{\pi}{2},\quad
\beta_0=\frac{\pi}{4},\quad \beta_1=-\frac{\pi}{4}
\label{eq:angles}
\end{equation}
give
\begin{equation}
S(\phi)=2\sqrt2\cos\phi,
\qquad S'(\phi)=-2\sqrt2\sin\phi.
\label{eq:Sphi}
\end{equation}
Away from isolated zero-probability boundary points, each term of Eq.~(\ref{eq:exactCorrFI}) equals unity; the same value is obtained at the boundaries by the regular limiting procedure. Therefore, for any setting allocation
\begin{equation}
\FC(\phi)=1.
\label{eq:FCideal}
\end{equation}
The QFI of $|\psi_\phi\rangle$ for the local generator $G=\sigma_z/2$ is
\begin{equation}
\FQ=4\Var(G)=1,
\end{equation}
so the complete CHSH coincidence record attains the QFI for all $\phi$.

This fact must not be confused with the sensitivity of the \emph{compressed Bell score}. Using Eqs.~(\ref{eq:scoreCHSH}) and (\ref{eq:corrBound}),
\begin{align}
\LB_E(\phi)&=\sin^2\phi,\label{eq:LEideal}\\
\LB_S(\phi)&=\frac{\sin^2\phi}{1+\sin^2\phi}.\label{eq:LSideal}
\end{align}
At $\phi=0$, where $S=2\sqrt2$ is maximally violated, both derivative-based certificates vanish even though $\FC=\FQ=1$. This is not a contradiction: $S(\phi)$ is locally stationary at its maximum, while the four outcome distributions still change with $\phi$ in directions canceled by the CHSH linear combination.

\begin{figure}[t]
\includegraphics[width=\columnwidth]{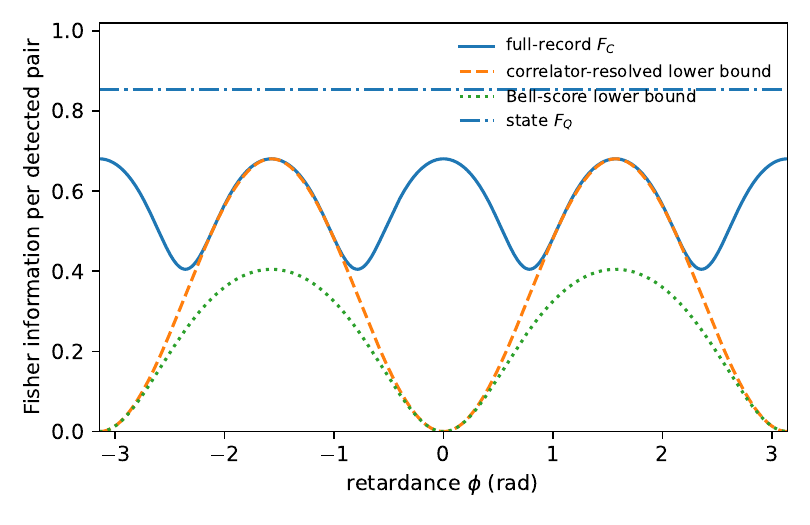}
\caption{Information hierarchy for canonical CHSH analyzers with depolarizing visibility $v=0.9$. The complete coincidence record carries more FI than either compressed certificate. The correlator-resolved certificate assumes unbiased marginals; the Bell-score certificate $S'^2/(16-S^2)$ does not. The QFI of the corresponding Werner-type state is also shown.}
\label{fig:hierarchy}
\end{figure}

\section{Bell--metrology tradeoff}

The ideal model makes the tension between Bell violation and score-based sensitivity analytic. Eliminating $\phi$ between Eqs.~(\ref{eq:Sphi}), (\ref{eq:LEideal}), and (\ref{eq:LSideal}) gives
\begin{align}
\LB_E&=1-\frac{S^2}{8},\label{eq:tradeE}\\
\LB_S&=\frac{8-S^2}{16-S^2},\label{eq:tradeS}
\end{align}
for $0\le |S|\le2\sqrt2$ along the canonical phase scan. CHSH violation requires $|S|>2$, and therefore in the violating region
\begin{equation}
\LB_E<\frac12,\qquad \LB_S<\frac13,
\end{equation}
with equality approached at the local bound $|S|=2$. At the Tsirelson point $|S|=2\sqrt2$, both derivative certificates vanish.

\begin{figure}[t]
\includegraphics[width=\columnwidth]{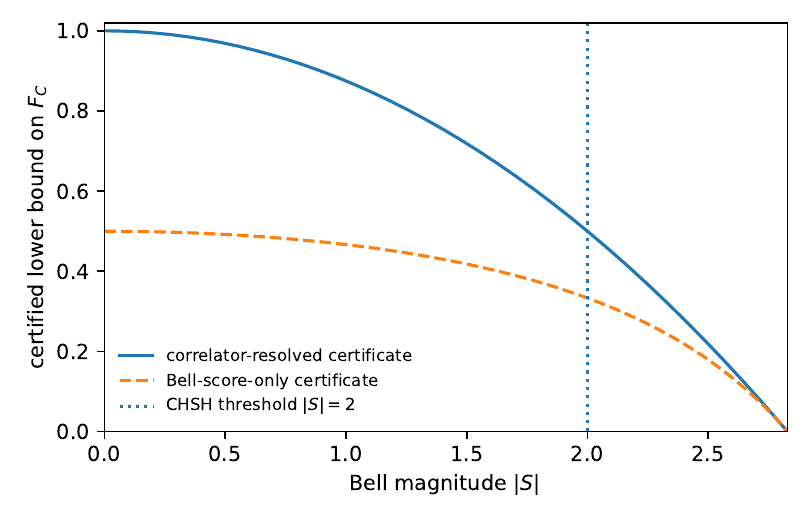}
\caption{Bell--metrology tradeoff for the ideal canonical scan. The vertical line marks the CHSH threshold. The complete-record FI is unity everywhere and is not shown. Strong Bell violation and a large first-order certificate from the scalar Bell score cannot be simultaneously maximized at the same working point.}
\label{fig:tradeoff}
\end{figure}

Figure~\ref{fig:tradeoff} shows the resulting Pareto structure. The conclusion is not that Bell violation is detrimental to metrology. The complete measurement remains QFI-optimal in the ideal model. Rather, a \emph{single scalar Bell score} is a lossy metrological compression, and the phase point that maximizes its violation is not the point that maximizes its slope-based sensitivity certificate.

\section{Separable benchmark and meaning of quantum advantage}

For the local encoding used above, entanglement does not increase the ultimate QFI if the fair benchmark allows an optimized pure separable probe. A resource-matched two-qubit separable probe can be prepared as
\begin{equation}
|+\rangle_A\otimes|\chi\rangle_B,
\qquad
|+\rangle_A=\frac{|H\rangle+|V\rangle}{\sqrt2},
\end{equation}
with the same $U_A(\phi)$ applied only to Alice; the arbitrary ancilla state $|\chi\rangle_B$ need not interact with the sample. This probe attains
\begin{equation}
\FQ^{\rm sep}=4\Var_{|+\rangle}(\sigma_z/2)=1.
\end{equation}
A suitable equatorial projective measurement attains this FI locally (up to the usual removable singular descriptions at fringe extrema). Hence, for this particular resource count (one encoded photon and, if desired, one unencoded ancilla photon per trial),
\begin{equation}
\FQ^{\rm Bell}=\FQ^{\rm sep}=1.
\end{equation}

This observation changes the correct interpretation of the protocol. The advantage demonstrated here is not an unconditional reduction of the Cram\'er--Rao limit relative to every optimized separable strategy. It is a \emph{dual-use certification advantage}: a data stream already collected for Bell characterization also provides a rigorous sensitivity statement, without complete state tomography. Other resource constraints can produce different comparisons. For example, comparing an entangled probe with its locally maximally mixed reduced state, or introducing correlated noise and reference-channel constraints, leads to a different benchmark and can favor nonlocal strategies, as explored experimentally and theoretically in Ref.~\cite{Pedram2024}. Such claims should be distinguished explicitly from the equal-encoded-photon benchmark considered here.

\section{Visibility loss, multipair emissions, and accidental coincidences}

\subsection{Symmetric background model}

A standard first model for imperfect entanglement is
\begin{equation}
\rho_v(\phi)=v|\psi_\phi\rangle\langle\psi_\phi|+(1-v)\frac{I}{4}.
\label{eq:Werner}
\end{equation}
Then
\begin{equation}
E_{xy}=v\cos(\alpha_x-\beta_y+\phi),\qquad
S=2\sqrt2v\cos\phi.
\end{equation}
The state QFI is
\begin{equation}
\FQ(v)=\frac{2v^2}{1+v},
\label{eq:QFIWerner}
\end{equation}
whereas the complete-record FI of the fixed canonical CHSH measurement is
\begin{equation}
\FC=\frac14\sum_{xy}
\frac{v^2\sin^2(\alpha_x-\beta_y+\phi)}
{1-v^2\cos^2(\alpha_x-\beta_y+\phi)}.
\label{eq:FCv}
\end{equation}
The two lower bounds become
\begin{align}
\LB_E(\phi)&=\frac{v^2\sin^2\phi}{2-v^2},\label{eq:LEv}\\
\LB_S(\phi)&=\frac{v^2\sin^2\phi}{2-v^2\cos^2\phi}.
\label{eq:LSv}
\end{align}
At the quadrature point $\phi=\pi/2$, the correlator-resolved bound saturates the fixed-measurement FI, while the score-only bound remains more conservative.

\begin{figure}[t]
\includegraphics[width=\columnwidth]{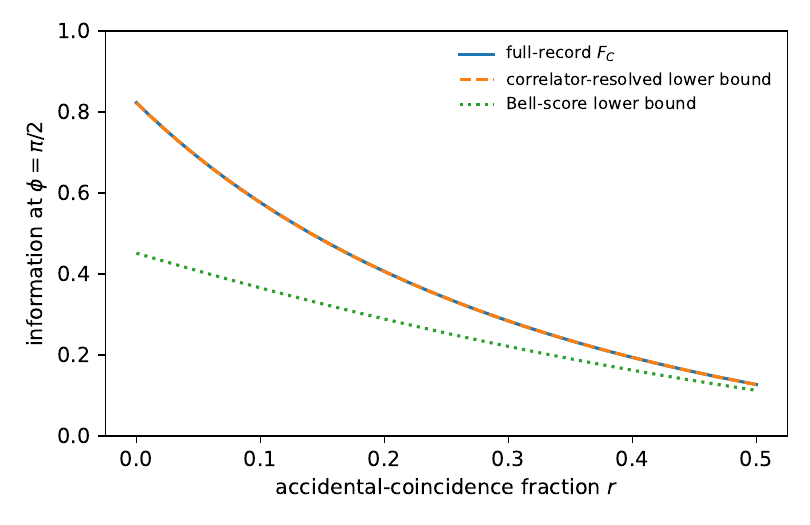}
\caption{Effect of symmetric accidental coincidences for a source visibility $v_0=0.95$ at $\phi=\pi/2$. A uniform accidental fraction $r$ produces $v_{\rm eff}=(1-r)v_0$. The bounds remain valid but become progressively weaker as the parameter-dependent coincidence contrast is diluted.}
\label{fig:acc}
\end{figure}

Continuous-wave and pulsed SPDC sources also generate multipair events. In a low-gain Poisson approximation with mean generated-pair number $\mu$ per effective coincidence window, $P(n\ge2)=1-e^{-\mu}(1+\mu)\simeq\mu^2/2$; multimode structure can change the detailed pair-number statistics, and the conversion from pair number to accidental coincidences also depends on efficiencies, dead time, and detector architecture. For this reason, a universal first-principles mapping from $\mu$ to the observed contamination fraction is not assumed here. After temporal filtering, unresolved independent pairs and detector dark/background events often contribute an approximately uncorrelated accidental component whose fraction can be measured from delayed time windows. A directly measurable phenomenological model is
\begin{equation}
P_{\rm obs}=(1-r_{\rm acc})P_{\rm sig}+r_{\rm acc}\frac14,
\label{eq:acc}
\end{equation}
where $r_{\rm acc}$ is the fraction of coincidences attributed to a uniform background, obtainable for example from shifted time windows. If the signal already has visibility $v_0$, Eq.~(\ref{eq:acc}) simply gives
\begin{equation}
v_{\rm eff}=(1-r_{\rm acc})v_0.
\end{equation}
Figure~\ref{fig:acc} shows the consequent monotonic loss of FI and both certificates at quadrature. Importantly, accidental coincidences do not invalidate the certificate; they reduce the measurable derivative and therefore reduce the certified information.

\subsection{Biased accidental background}

Accidentals need not be uniform. A simple factorized background is
\begin{equation}
P_{\rm acc}(a,b|x,y)=\frac{(1+a u_x)(1+b w_y)}{4},
\end{equation}
which can describe unequal singles rates or efficiency imbalance. With
\begin{equation}
P_{\rm obs}=(1-r)P_{\rm sig}+rP_{\rm acc},
\end{equation}
one obtains nonzero marginals $m_x=ru_x$ and $n_y=rw_y$. The unbiased identity Eq.~(\ref{eq:exactCorrFI}) and its correlator-resolved bound should then not be used. The correct FI is Eq.~(\ref{eq:fullbiasedFI}), while the Bell-score bound Eq.~(\ref{eq:generalBellBound}) remains valid without modification. This is one reason the score-only theorem is useful despite being looser: it survives exactly the detector and source asymmetries that invalidate the compact correlator formula.

\section{Finite-count certification without ratio ambiguities}

A finite-sample statement should bound the \emph{final ratio} on a simultaneous high-probability event, rather than combine unrelated point estimates. We give such a construction for CHSH. Let $\widehat E_{xy}$ be the mean of $t_i=a_ib_i\in\{\pm1\}$ from $N_{xy}$ events. Hoeffding's inequality \cite{Hoeffding1963} gives
\begin{equation}
\Pr\left(|\widehat E_{xy}-E_{xy}|>\epsilon_{xy}\right)\le \alpha_{xy},
\qquad
\epsilon_{xy}=\sqrt{\frac{2}{N_{xy}}\ln\frac{2}{\alpha_{xy}}}.
\label{eq:hoeff}
\end{equation}
Therefore, at a fixed phase,
\begin{equation}
|\widehat S-S|\le \Delta_S\equiv\sum_{xy}\epsilon_{xy}
\label{eq:Srad}
\end{equation}
whenever all four correlator events hold. A union bound controls all required phases simultaneously and does not require statistical independence between the resulting confidence events.

Suppose the retardance is probed at $\phi_0$ and $\phi_0\pm\delta$. For the sinusoidal canonical model, including a constant background offset, the central derivative obeys the exact identity
\begin{equation}
S'(\phi_0)=\frac{S(\phi_0+\delta)-S(\phi_0-\delta)}{2\sin\delta}.
\label{eq:exactdiff}
\end{equation}
Hence
\begin{equation}
L_{S'}=\max\left\{0,
\left|\frac{\widehat S_+-\widehat S_-}{2\sin\delta}\right|
-\frac{\Delta_{S,+}+\Delta_{S,-}}{2\sin\delta}
\right\}
\label{eq:Lsprime}
\end{equation}
provides a simultaneous lower bound on $|S'|$. At the working point,
\begin{equation}
L_S=\max\{0,|\widehat S_0|-\Delta_{S,0}\}
\end{equation}
provides a lower bound on $|S|$. Since the denominator in Eq.~(\ref{eq:scoreCHSH}) is $16-S^2$, an upper bound on the denominator is $16-L_S^2$. Therefore, on the same simultaneous confidence event,
\begin{equation}
\boxed{
\FC(\phi_0)\ge
\underline F_C\equiv
\frac{L_{S'}^2}{16-L_S^2}.
}
\label{eq:finiteLCB}
\end{equation}
This construction is mathematically monotone: the numerator is replaced by a simultaneous lower bound and the positive denominator by a simultaneous upper bound. No assumption of independent numerator and denominator error bars is used.

For a general smooth $S(\phi)$ rather than the sinusoidal optical model, one may replace $2\sin\delta$ by $2\delta$ and subtract the deterministic central-difference remainder $M_3\delta^2/6$ from $L_{S'}$ if $|S^{(3)}|\le M_3$ is known over the modulation interval. Tighter empirical-Bernstein radii can replace Eq.~(\ref{eq:hoeff}) \cite{MaurerPontil2009}.

\begin{figure}[t]
\includegraphics[width=\columnwidth]{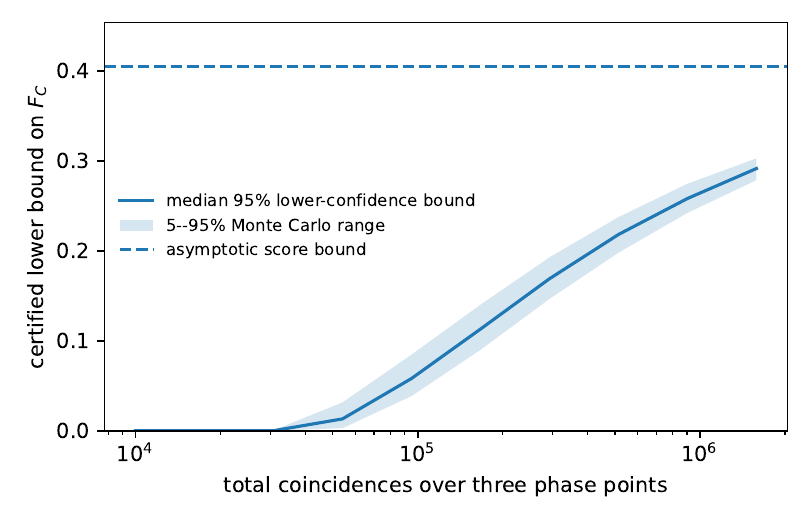}
\caption{Finite-count certification of the Bell-score bound for $v=0.9$, $\phi_0=\pi/2$, and $\delta=0.1$ rad. Twelve correlators are used (four at each of three phase points), with equal allocation and a $95\%$ simultaneous Hoeffding construction. For each count budget, 250 Monte Carlo realizations were generated with seed 20261011. The solid curve is the median lower-confidence bound; the band shows the 5--95 percentile range; the horizontal line is the asymptotic score certificate.}
\label{fig:finite}
\end{figure}

Figure~\ref{fig:finite} illustrates Eq.~(\ref{eq:finiteLCB}) for $v=0.9$, $\phi_0=\pi/2$, $\delta=0.1$ rad, equal setting allocation, and joint confidence $95\%$. For each coincidence budget we generated 250 Monte Carlo realizations with pseudorandom seed 20261011. The total budget is divided equally among the three phase points and then among the four CHSH setting pairs. Twelve correlator intervals are combined with a Bonferroni/union-bound allocation $\alpha=0.05/12$. The lower bound converges toward the asymptotic score certificate as the coincidence budget grows. This score-level certificate is conservative by design and remains valid even if the single-party marginals are biased.

\section{Measurement economy versus tomography}

The proposed hierarchy clarifies the comparison with full quantum-state tomography (QST). QST can reconstruct a model of the density matrix and, after specifying the encoding generator, can be used to calculate QFI or optimize a measurement. That additional information can produce a tighter characterization, but it requires an informationally complete set of measurement configurations and introduces reconstruction assumptions and error propagation.

The Bell-score certificate uses less information on purpose. The scalar pair $(S,S')$ gives the robust bound Eq.~(\ref{eq:scoreCHSH}); the four correlators and their derivatives give the tighter Eq.~(\ref{eq:corrBound}) when the marginals are balanced; and the complete outcome frequencies give Eq.~(\ref{eq:fullbiasedFI}) without reconstructing the underlying density matrix. The price of compression can be quantified in the ideal model. At quadrature, $\phi=\pi/2$, the full CHSH record has $\FC=1$, the correlator-resolved certificate also gives $\LB_E=1$, whereas the Bell-score certificate gives only $\LB_S=1/2$. At the point of maximal Bell violation, $\phi=0$, the full record still has $\FC=1$ but both derivative-based compressed certificates vanish. For $v=0.9$ at quadrature, the fixed canonical measurement gives $\FC=\LB_E=v^2/(2-v^2)\simeq0.681$, the score-only certificate gives $\LB_S=v^2/2\simeq0.405$, and the Werner-state QFI is $\FQ=2v^2/(1+v)\simeq0.853$.

These numbers make the comparison with QST more concrete. Tomography may support a tighter state-level characterization and subsequent measurement optimization, but it requires an informationally complete data set and propagates reconstruction uncertainty. The CHSH route deliberately accepts a looser statement in exchange for using a much smaller, already familiar measurement set. The tradeoff is therefore
\begin{equation}
\text{measurement economy}\ \leftrightarrow\ \text{characterization tightness},
\end{equation}
not an assertion that Bell statistics replace complete state information.

\section{Experimental protocol}

A minimal implementation uses a polarization-entangled SPDC source, a birefringent sample or tunable retarder in Alice's arm, and two polarization-analyzer settings on each side. A practical protocol is:

\begin{enumerate}
\item Randomize the four CHSH setting pairs with known parameter-independent probabilities $p_{xy}$ (uniform for standard CHSH).
\item Record the four coincidence outcomes rather than only the correlator sums; this preserves the option to diagnose biased marginals and evaluate Eq.~(\ref{eq:fullbiasedFI}).
\item Acquire data at the target retardance and under a small calibrated modulation $\phi_0\pm\delta$.
\item Report $S$, the single-party marginals, and the full outcome frequencies. If marginal confidence intervals are compatible with zero, report both $\LB_E$ and $\LB_S$; otherwise report the full-record FI from a calibrated likelihood together with the assumption-robust $\LB_S$.
\item Estimate accidental coincidences from delayed coincidence windows or another independent timing calibration and report their fraction. The bound should be applied to the observed probabilities; background subtraction is not required for validity and, if used, must propagate its uncertainty.
\item For finite counts, construct the simultaneous lower-confidence bound in Eq.~(\ref{eq:finiteLCB}) rather than inserting unconstrained point estimates into the asymptotic ratio.
\end{enumerate}

This workflow produces three scientifically distinct outputs from the same measurement stream: a Bell statistic, a fixed-measurement FI (when the likelihood response is calibrated), and a guaranteed Bell-statistic lower bound on that FI.

\section{Discussion}

The main lesson is straightforward: a Bell number and a Fisher information are different summaries of the same underlying probability model. The Bell value tests a constraint on correlations; the FI measures how the probabilities move when the parameter changes. What links them is the derivative of the statistic, not the Bell violation itself. Equation~(\ref{eq:generalBellBound}) is simply the information inequality applied to that parameter-dependent Bell statistic.

The novelty is therefore not the first-moment information inequality by itself. Related metrological-witness approaches lower-bound QFI from incomplete observables \cite{Apellaniz2017}, moment methods lower-bound FI from partial distributional information \cite{Len2022}, and Bell correlations have previously been linked to many-body quantum-enhanced metrology \cite{Niezgoda2021}. Here the object being certified is instead the FI of a specified Bell-experiment record. This distinction permits a closed CHSH score bound that remains valid with biased marginals, makes the information loss under Bell-score compression explicit, and supports a nonasymptotic lower-confidence certificate directly from finite coincidence counts.

This distinction separates three questions that are easy to conflate. A Bell violation asks whether $B$ crosses a local bound. Metrological sensitivity asks how rapidly the outcome distribution changes and is quantified by FI. Quantum advantage asks whether that sensitivity beats the best strategy allowed under the same resource accounting. The retardance model provides a useful example because these three questions give different answers at the same operating point.

At maximal CHSH violation the scalar Bell score has zero slope, but the individual coincidence probabilities still move with $\phi$. The loss therefore comes from compression, not from an absence of information in the experiment. Keeping the four correlators recovers more of the available sensitivity, and keeping the full outcome probabilities is the appropriate route when detector imbalance or other effects bias the marginals. This also explains why tomography can give a tighter state-level assessment: it retains still more information, at the price of a larger measurement and reconstruction problem.

The same statistical construction is not limited to CHSH. Any linear statistic built from discrete outcomes can be treated in the same way once its event-level random variable is specified. Tilted-CHSH inequalities, multiparty full-correlation expressions, and steering functionals under the appropriate trust model are natural extensions. In a multiparameter problem, the derivative becomes a gradient and the corresponding statement is a matrix inequality involving the Fisher information matrix.

A useful experimental test would deliberately vary both accidental background and detector imbalance while recording the full coincidence table. One could then compare the score-only certificate, the correlator-resolved certificate, and the FI of the full likelihood on the same data. Such an experiment would show directly where the balanced-marginal approximation breaks down and how many detected events are needed before the finite-count certificate becomes informative.

\section{Conclusion}

We have recast the connection between Bell measurements and retardance sensing as a statistical-information problem. For a differentiable full-correlation Bell score, its parameter slope gives a lower bound on the Fisher information of the complete observed record. In the uniformly sampled CHSH case the bound is $S'^2/(16-S^2)$ and remains valid even when the single-party marginals are biased. When the marginals are balanced, the four correlators carry enough information to give a tighter certificate and an exact FI for the chosen measurement.

The polarization example shows why that hierarchy matters. The CHSH value is largest at a point where its own first derivative is zero, so the Bell-score certificate vanishes even though the full coincidence record remains sensitive to retardance. An optimized separable probe can also reach the same QFI for this local encoding, which rules out a blanket claim of entanglement-enhanced precision. What the protocol offers instead is a practical dual use of Bell-calibration data: the same measurement record can support a nonlocality analysis and a controlled statement about parameter sensitivity.

Biased marginals, accidental coincidences, and finite counts can all be included without changing that basic interpretation. The result is deliberately modest in scope, but stronger for being so: Bell data do not determine the Fisher information, yet their parameter response can place a rigorous lower bound on it.

\section*{Acknowledgements}
The authors acknowledge the Universidad Aut\'onoma del Estado de M\'exico for institutional support.

\section*{Funding}
This research received institutional support from the Universidad Aut\'onoma del Estado de M\'exico. No specific grant number is associated with this work.

\section*{Conflict of Interest}
The authors declare no conflict of interest.

\section*{Data Availability Statement}
The data underlying the analytical and numerical results are generated by the equations and simulation procedures described in the article. Python scripts used for the numerical checks and figure generation are included as ancillary files with this preprint.

\section*{Generative AI Disclosure}
OpenAI ChatGPT was used as an assistive tool during manuscript development to help reorganize text, check the internal consistency of algebraic derivations, and improve language and presentation. All equations, numerical results, references, interpretations, and final wording were independently reviewed and approved by the authors, who take full responsibility for the content.

\appendix

\section{Proof of the score theorem in likelihood notation}

Let $z=(x,y,a,b)$ and $q_\phi(z)$ be a differentiable discrete distribution with parameter-independent support. For any parameter-independent event statistic $W(z)$,
\begin{align}
\partial_\phi\E[W]
&=\sum_z W(z)\partial_\phi q_\phi(z)\\
&=\sum_z W(z)q_\phi(z)\partial_\phi\ln q_\phi(z)\\
&=\E[W\ell_\phi].
\end{align}
Normalization gives $\E[\ell_\phi]=0$, hence $\partial_\phi\E[W]=\operatorname{Cov}(W,\ell_\phi)$. Cauchy--Schwarz gives
\begin{equation}
|\partial_\phi\E[W]|^2\le \Var(W)\Var(\ell_\phi)
=\Var(W)\FC.
\end{equation}
Using Eq.~(\ref{eq:W}), $\E[W]=B$ and $\E[W^2]=\sum c_{xy}^2/p_{xy}$, proving Eq.~(\ref{eq:generalBellBound}).

\section{Derivation of the unbiased correlator FI}

For a fixed setting pair write $E=E_{xy}$ and
\begin{equation}
P_{ab}=\frac{1+abE}{4},\qquad
\partial_\phi P_{ab}=\frac{abE'}{4}.
\end{equation}
Then
\begin{align}
F_{xy}
&=\sum_{ab}\frac{(\partial_\phi P_{ab})^2}{P_{ab}}\\
&=\frac{E'^2}{4}\sum_{ab}\frac{1}{1+abE}\\
&=\frac{E'^2}{1-E^2}.
\end{align}
Summing with weights $p_{xy}$ gives Eq.~(\ref{eq:exactCorrFI}).

\section{Explicit polarization-qubit derivation}\label{app:qubitderivation}

With $|0\rangle\equiv|H\rangle$ and $|1\rangle\equiv|V\rangle$,
\begin{equation}
|\psi_\phi\rangle=\frac{e^{-i\phi/2}|00\rangle+e^{i\phi/2}|11\rangle}{\sqrt2}.
\end{equation}
For $\sigma_\theta=\cos\theta\,\sigma_x+\sin\theta\,\sigma_y$,
\begin{equation}
\sigma_\theta|0\rangle=e^{i\theta}|1\rangle,\qquad
\sigma_\theta|1\rangle=e^{-i\theta}|0\rangle.
\end{equation}
Using the analyzer convention $A_x=\sigma_{-\alpha_x}$ and $B_y=\sigma_{\beta_y}$ gives
\begin{align}
E_{xy}(\phi)
&=\langle\psi_\phi|A_x\otimes B_y|\psi_\phi\rangle\\
&=\cos(\phi+\alpha_x-\beta_y),
\end{align}
which is Eq.~(\ref{eq:Eideal}). Substituting the angles in Eq.~(\ref{eq:angles}) yields
\begin{align}
E_{00}&=\cos(\phi-\pi/4),&
E_{01}&=\cos(\phi+\pi/4),\\
E_{10}&=\cos(\phi+\pi/4),&
E_{11}&=\cos(\phi+3\pi/4).
\end{align}
Therefore
\begin{align}
S(\phi)&=E_{00}+E_{01}+E_{10}-E_{11}\\
&=2\sqrt2\cos\phi,
\end{align}
and differentiation gives $S'(\phi)=-2\sqrt2\sin\phi$. This derivation also makes clear that the sign convention for the analyzer phase labels has no physical effect; it only fixes the compact form of Eq.~(\ref{eq:Eideal}).

\section{Werner-state QFI}

For Eq.~(\ref{eq:Werner}), the eigenvalue of $|\psi_\phi\rangle$ is $\lambda_1=(1+3v)/4$ and the three orthogonal eigenvalues are $\lambda_0=(1-v)/4$. The local generator $G=\sigma_z/2$ couples the Bell component only to its orthogonal phase-flipped Bell partner, with squared matrix element $1/4$. The standard spectral formula for unitary QFI therefore gives
\begin{equation}
\FQ=4\frac{(\lambda_1-\lambda_0)^2}{\lambda_1+\lambda_0}\frac14
=\frac{2v^2}{1+v}.
\end{equation}

\section{General smooth finite-difference certificate}

If $S$ is three times differentiable and $|S^{(3)}(\phi)|\le M_3$ for $\phi\in[\phi_0-\delta,\phi_0+\delta]$, then
\begin{equation}
\left|S'(\phi_0)-\frac{S(\phi_0+\delta)-S(\phi_0-\delta)}{2\delta}\right|
\le \frac{M_3\delta^2}{6}.
\end{equation}
Consequently Eq.~(\ref{eq:Lsprime}) is replaced by
\begin{equation}
L_{S'}=\max\left\{0,
\left|\frac{\widehat S_+-\widehat S_-}{2\delta}\right|
-\frac{\Delta_{S,+}+\Delta_{S,-}}{2\delta}
-\frac{M_3\delta^2}{6}
\right\}.
\end{equation}
The same denominator construction then proves Eq.~(\ref{eq:finiteLCB}) on the joint confidence event.

\end{document}